\documentstyle[12pt]{article}
\begin{document}
\title{ \vspace{-10mm}
       {\normalsize
       \begin{tabbing}
       \'March 1993   \`hep-th/9304096
    \\  \' DESY 93-051
       \end{tabbing} }
       \vspace{10mm}
A Monopole Solution in open String Theory}

\author{K. Behrndt\thanks{e-mail: behrndt@r6rz.ifh.de}\\
{\normalsize \em DESY-Institut f\"ur Hochenergiephysik, Zeuthen}}
\date{}
\maketitle
\begin{abstract}
We investigate a solution  of the Weyl invariance conditions
in open string theory in 4 dimensions.
In the closed string sector this solution
is a combination of the SU(2) Wess--Zumino--Witten model and a Liouville
theory. The investigation is carried out in the $\sigma$ model approach
where we have coupled all massless modes (especially an abelian gauge
field via the boundary) and tachyon fields. Neglecting all higher
derivatives in the field strength we get an exact result which
can be interpreted as a monopole configuration living in non--trivial
space time. The masses of both tachyon fields are quantized by $c_{wzw}$
and vanish for $c_{wzw} = 1$.
\end{abstract}

\renewcommand{\arraystretch}{1.6}
\renewcommand{\thefootnote}{\alph{footnote}}
\newcommand{\pa}{\partial}
\newcommand{\be}{\begin{equation}}
\newcommand{\ee}{\end{equation}}
\newcommand{\ba}[1]{\begin{array}{#1}}
\newcommand{\ea}{\end{array}}
\newcommand{\aap}{\alpha^{\prime}}
\newcommand{\si}{\mbox{$\sigma$}~}
\newcommand{\bb}{\mbox{$\beta$}}
\newcommand{\bbb}{\mbox{$\bar{\beta}$}}
\newcommand{\vsf}{\vspace{3mm}}
\section{Introduction}
Solutions of the conformal invariance conditions in closed string theory
have been widely discussed in the last time. First of all these are
solutions corresponding to (gauged) Wess--Zumino--Witten (WZW)
theories (see e.g.~\cite{lust}). It is possible to obtain  a non-linear
\si model by starting with a gauged WZW theory, choosing a gauge and
integrating out the $2d$ gauge field. From this \si model one can read off
the metric, antisymmetric tensor and the dilaton. Usually one gets
by this approach the background field only in an $\aap$ expansion.
In order to get exact results one has to use algebraic Hamiltonian techniques
\cite{verl} or to compute first the effective action of the gauged WZW model
and then to eliminate the $2d$ gauge field \cite{bars}. Then the corresponding
background fields solve the $\bbb$ equations
in all orders in $\aap$. Another class  are the instanton,
soliton and monopole solutions \cite{strom,khur} that one gets from
the heterotic string
theory with self dual non-abelian gauge field. They have been constructed
in the lowest order in $\aap$ and are exact due to the non--renormalization
theorem in extended worldsheet supersymmetric theories \cite{howe}. The
soliton solution corresponds
to a semi-wormhole in the space time and near the singularity this theory
is approximately given by a combination of a SU(2) WZW model and a
Liouville theory (after incorporation of a tachyon field). Therefore, in
this limit it is also possible to construct an exact conformal theory
in the bosonic case. The generalization of
this closed string theory to an open string theory is the aim
this paper.

\vsf

The \si model containing the massless modes and tachyon field of
the closed and open string is given by
\begin{equation}
  \begin{array}{rcl}
    Z&=&\int D x ~ e^{-S} ~~, \\
    S&=&\frac{1}{4\pi\aap} \int_M d^2 z \sqrt{g}\left(g^{ab}
     \partial_a x^{\mu} \partial_b x^{\nu} G_{\mu\nu} -
  i\frac{\epsilon^{ab}}{\sqrt{g}} \partial_a x^{\mu} \partial_b
    x^{\nu}  B_{\mu\nu} + \alpha' R^{(2)} \phi
     + \alpha' T_1 \right) +   \\
   && + \frac{1}{2\pi\alpha'}\int_{\pa M} \left( \aap \,k \,\phi + \aap\,
       T_2 + i\, e A_{\mu} \, \dot{x}^{\mu} \right)~.
  \end{array}
\end{equation}
Here $G_{\mu\nu}$ corresponds to the metric in the target space
(space time), $B_{\mu\nu}$ is the antisymmetric tensor field, $\phi$
is the dilaton, $A_{\mu}$ is an abelian gauge field, $T_{1,2}$
the tachyon fields for closed or open string respectively. We have to
decide between both tachyons because they carry different world sheet
dimensions. $R^{(2)}$ is the worldsheet scalar curvature, $k$  the
corresponding curvature of the boundary and $e$ an electric charge. This
model is conformally invariant if the Weyl anomaly coefficients vanish.
These coefficients are the \bbb~functions which are mainly given by the
renormalization group \bb~functions \cite{tseyt2}.
In general it is only possible to solve these equations in an
$\aap$ expansion. But for an appropriate choice of the background fields
it is possible to find exact solutions, e.g.~as a WZW theory.

Starting with an exact solution in the closed string sector (sect.~2)
we will show in sect.~3 that it is possible to find an exact solution
for the \bbb~equations in the open string sector too. Both
sections are rather technical whereas in section 4 we
summarize all results and give an interpretation and discussion of the
results.
\section{Closed string sector}
We start with the investigation of the closed string sector
and thereafter we study the generalization to open strings.
So we neglect here the couplings via the boundary in (1).
As we have already noted for special background fields and in four space
time dimensions
it is possible to rewrite the model (1) in a $SU(2)$ WZW model and
a Liouville model \cite{khur,elis,behr}. Namely, if
\begin{equation}
\begin{array}{l}
G_{\mu\nu}=\frac{Q}{r^2}\delta_{\mu\nu} \qquad \qquad \mu,\nu = 0,1,2,3,\\
H_{\mu\nu\lambda} \equiv \pa_{[\mu}B_{\nu\lambda]} = \pm
  \epsilon_{\mu\nu\lambda}^{~~~~\sigma}
  \partial_{\sigma}\log\frac{\sqrt{Q}}{r}\\
\phi = \phi(r) \quad \mbox{and}\quad T_1=T_1(r)
\end{array}
\end{equation}
we can write (1) as
\begin{equation}
\begin{array}{l}
S = S_{WZW}(SU(2)) + \frac{1}{4\pi\alpha'}\int\left( (\partial u)^2 Q
    + \alpha' R^{(2)} \phi(u) + \alpha' T_1(u) \right)\\
\mbox{and}\\
S_{WZW} = \frac{Q}{4\pi\alpha'}\int d^2 z \, tr\left( \partial_a g^{-1}
   \partial_a g \right) + \frac{ Q}{6 \pi\alpha'}\int d^3 z \, tr\,
   \epsilon^{abc} \left( (g^{-1}\partial_a g) (g^{-1} \partial_b g)
  (g^{-1} \partial_c   g) \right)
\end{array}
\end{equation}
where $g = \frac{1}{r}(x^0 {\bf 1} + i x^{i}${\bf $\sigma^i$}),
$\sigma^i$ are the Pauli matrices, $u=\log r$, $r^2=(x^0)^2+(x^1)^2+(x^2)^2+
(x^3)^2$ and $Q = const.$ After introducing polar coordinates one
finds immediately that the angle degrees of freedom are controlled
by the WZW part and the second Liouville part contains the
dependence on the radius. Background fields like (2)
have been investigated in the context of semi--wormholes
or solitons in \cite{strom} and its cosmological relevance is discussed in
\cite{elis}.
The WZW theory is well defined and conformally invariant if \cite{witt2}:
$\frac{Q}{\aap} \equiv k = 1,2,3,...$ (the level of the WZW theory).
The corresponding central charge for the $SU(2)$ is given by:
$c_{wzw}=\frac{3k}{k+2}$. In addition, the conformal invariance
requires that the tachyon and dilaton have to fulfill
the \bbb~equations of the Liouville part in (3)
\footnote{We neglect in
our consideration all non-perturbative contributions, e.g.
a tachyon potential.}
\begin{equation}
\begin{array}{c}
\phi''(u)=0\\
1+c_{wzw}-26 + \frac{6}{k}\, (\phi ')^2=0\\
-\frac{1}{2 k}\, T_1'' - 2T_1 + \frac{1}{k} \, \phi' T_1' = 0
\end{array}
\end{equation}
(26 is the contribution of the ghosts to the central charge and
$\phi'(u) \equiv \frac{d}{du} \phi(u)$). From the first equation it follows
that the dilaton is at most linear in $u$, the second equation ensures
the vanishing of the total central charge and the last one defines the
tachyon field. As solution one finds
\begin{equation}
\begin{array}{l}
\phi(u)=\phi_0 + q u \qquad \mbox{with}: \qquad
 q^2 = \frac{k}{6}(25 - \frac{3k}{k+2}) ~, ~ u=\log r ~,\\
 T_1^{(1)}(u) \sim e^{q u} \sqrt{\frac{k}{6}(\frac{3k}{k+2}-1)}^{\,-1}
 \sin (\sqrt{\frac{k}{6}(\frac{3k}{k+2}-1)}u) \quad ,\\
 T_1^{(2)}(u) \sim e^{q u} \cos(\sqrt{\frac{k}{6}(\frac{3k}{k+2}-1)} u)~.
\end{array}
\end{equation}
These exact results have been obtained by rewriting the original
\si model in the known WZW model and the Liouville theory. Since this
procedure is not practicable in open string theory we want now
to consider this result from the \si model point of view.
This means we have to show that the background fields (2) and (5)
correspond to zeros of the \bbb~functions and therewith
to a conformal field theory. These functions are given by \cite{tseyt2}
\begin{equation}
 \begin{array}{l}
 \bar{\beta}^{G}_{\mu\nu} =
                  \beta^{G}_{\mu\nu} + D_{(\mu}M_{\nu)}
                 \qquad ,\qquad \qquad  M_{\nu} = 2\alpha'
                \partial_{\nu} \phi + W_{\nu}~,\\
   \bar{\beta}^{B}_{\mu\nu}= \beta^B_{\mu\nu} + H_{\mu\nu\lambda}
       M^{\lambda} + \partial_{[\mu} K_{\nu]}~,\\
    \bar{\beta}^{\phi} = \beta^{\phi} +
                          \frac{1}{2} M^{\mu}\partial_{\mu}\phi~, \\
  \bar{\beta}^T_1 = \beta^T_1 - 2T_1 + \frac{1}{2} M^{\mu}\partial_{\mu}T_1 ~.
 \end{array}
\end{equation}
Up to the second order in $\alpha'$ one gets for the renormalization
group $\beta$ functions
\begin{equation}
 \begin{array}{rcl}
  \beta^{T_{1}}&=&-\frac{1}{2} \alpha' D^2 T_1 - \alpha'^2\,\frac{1}{8}
      (H^2)^{\mu\nu} D_{\mu}\partial_{\nu}T_1 ~, \\
  \beta^{G}_{\mu\nu}&=&\alpha' \hat{R}_{(\mu\nu)} + \frac{1}{2} \alpha'^2
      \left(\hat{R}^{\alpha\beta\lambda}_{~~~(\nu}\hat{R}_{\mu)\alpha
      \beta\lambda} -\frac{1}{2} \hat{R}^{\beta\lambda\alpha}_{~~~(\nu}
      \hat{R}_{\mu)\alpha\beta\lambda} +\frac{1}{2}
      \hat{R}_{\lambda(\mu\nu)\beta} (H^2)^{\lambda\beta}\right)~,\\
  \beta^{B}_{\mu\nu}&=&\alpha' \hat{R}_{[\mu\nu]} + \frac{1}{2} \alpha'^2
      \left(\hat{R}^{\alpha\beta\lambda}_{~~~[\nu}\hat{R}_{\mu]\alpha
      \beta\lambda} -\frac{1}{2} \hat{R}^{\beta\lambda\alpha}_{~~~[\nu}
      \hat{R}_{\mu]\alpha\beta\lambda} +\frac{1}{2}
      \hat{R}_{\lambda[\mu\nu]\beta} (H^2)^{\lambda\beta}\right)~,\\
  \beta^{\phi}&=&\frac{1}{6}(d-26) - \frac{1}{2}\alpha' D^2 \phi -
      \frac{1}{8} \alpha'^{2} (H^2)^{\mu\nu}D_{\mu}D_{\nu}
      \phi + \\
    & & +\frac{1}{16}\alpha'^2\left(R_{\mu\nu\alpha\lambda}^2 -
         \frac{11}{2} R H H +
      \frac{5}{24}H^4 + \frac{3}{8}(H^2_{\mu\nu})^2 + \frac{4}{3}
      D H \cdot D H  \right)
 \end{array}
\end{equation}
and $W_{\mu} = -(\alpha'^2 /24) \partial_{\mu}H^2$, $K_{\mu}={\cal O}
(\alpha'^3)$, $H_{\mu\nu\lambda}=\partial_{[\mu}B_{\nu\lambda]}$,
$D H \cdot D H \equiv D_{\mu}H_{\nu\lambda\beta} D^{\mu}
H^{\nu\lambda\beta}$
and $\hat{R}_{\mu\nu\lambda\beta}$ is the generalized curvature tensor
computed in terms of the connection $\hat{\Gamma}^{\mu}_{~\nu\lambda}
 = \Gamma^{\mu}_{~\nu\lambda} - \frac{1}{2} H^{\mu}_{~\nu\lambda}$
\be
\hat{R}_{\mu\nu\lambda\gamma}=R_{\mu\nu\lambda\gamma}+
   \frac{1}{2} \left( D_{\gamma} H_{\mu\nu\lambda} - D_{\lambda}
   H_{\mu\nu\gamma}\right) + \frac{1}{4}\left(H_{\mu\gamma\rho}
   H^{\rho}_{~\nu\lambda} - H_{\mu\lambda\rho}
   H^{\rho}_{~\nu\gamma}\right)~.
\ee
It is now easy to prove that $\hat{R}$ vanishes for
the metric and the torsion (2), i.e. the space time is
parallelizable \cite{brat}. The vanishing of $\hat{R}$ has the remarkable
consequence that the \bb~ functions of the metric and of
the antisymmetric tensor vanish identically in all
order in $\aap$ \cite{muki}. Furthermore, the torsion and the Riemann
curvature are covariantly constant: $D_{\lambda}H_{\mu\nu\gamma} \equiv 0$,
$D_{\lambda}R_{\mu\nu\rho\gamma} \equiv 0$.
Let us show that $W_{\nu}$ and $K_{\nu}$ do not contribute to the
$\bar{\beta}$ functions. Both expressions are functions of the
curvature and torsion only. Since $\hat{R}=0$, we can substitute
all curvature dependence by additional torsion terms. These terms
are covariantly constant and thus: $D_{\mu}W_{\nu}=D_{\mu}K_{\nu} =
0$. So, they do not contribute to $\bar{\beta}^{G,B}$.
In addition, $W_{\nu}$ contributes to $\bar{\beta}^{\phi ,T}$
in the form of $W^{\nu}\pa_{\nu}\phi \sim W_{\nu}x^{\nu}$ which vanish too
because $H_{\lambda\mu\nu}x^{\nu}=0$. The
tachyon term has the same structure because it depends on the radius
only. Hence we can always set $W_{\nu}=K_{\nu}=0$ or
$M_{\nu}=2\aap\pa_{\nu}\phi$.
The whole \bbb~functions of the metric and the antisymmetric
tensor are thus given by
\be
\ba{l}
\bbb^{G}_{\mu\nu} \sim D_{\mu}D_{\nu}\phi ~,\\
\bbb^{B}_{\mu\nu} \sim H_{\mu\nu}^{~~\lambda} \pa_{\lambda} \phi
\ea
\ee
and we find that for the dilaton (5) and the torsion (2) both expressions
vanish. The Curci--Paffuti theorem \cite{curc} has now the consequence that
the dilaton \bbb~function has to be constant, namely, the central charge.
Because the dilaton \bbb~function is only perturbatively known
one can determine the parameter $q$ in the \si~model approach only order
by order in $\aap$. Khuri \cite{khur} has shown that one just
gets the correct expansion in $\frac{\aap}{Q}=1/k$.
If the tachyon depends on the radius only it is also possible to obtain an
exact solution for the tachyon \bbb~equation. Although one does
not know $\bar{\beta}^{T_1}$ in all order of the $\aap$ expansion the
general structure is known \cite{tseyt1}
\be
 \bbb^{T_1} = P^{\mu\nu} D_{\mu}\pa_{\nu} T_1
 -2 T_1 + \aap \pa^{\mu}\phi \pa_{\mu} T_{1} ~.
\ee
Where $P^{\mu\nu}$ is a general operator depending on the curvature,
the torsion, the covariant derivative  and the metric \cite{tseyt1}.
In the flat limit where
all curvature and torsion terms drop out $P^{\mu\nu}$ is given by:
$P^{\mu\nu} = -\frac{1}{2}\aap G^{\mu\nu}$.
Since $\hat{R}_{\mu\nu\lambda\sigma} = 0$ and all covariant derivatives
of $R$ and $H$ vanish $P^{\mu\nu}$ can be written as a function of $H$ and
covariant derivatives acting on the tachyon.
Crucial for this argumentation is that: $D_{\mu}\pa_{\nu} T_1(r) \sim
\frac{x^{\mu}}{r^2}\,\frac{x^{\nu}}{r^2}$ and that $\frac{x^{\mu}}{r^2}$
is a covariantly constant vector which can be commutated with the covariant
derivatives. Since furthermore $H_{\mu\nu\lambda} x^{\lambda} = 0$ we find
that all curvature and torsion terms drop out of $P^{\mu\nu}$ and we get
the same structure as in the flat limit
\be
 \bbb^{T_1} = -\frac{1}{2} \aap D^2 T_1
   - 2 T_1 + \aap \pa^{\mu}\phi \pa_{\mu} T_1 ~.
\ee
It is easy to verify that for our solution (5) this expression vanish.

Similar to the WZW approach one obtains the quantization of
$k = \frac{Q}{\aap}$ only via a topological consistency condition.
Namely, if $g$ is an element of the SU(2)
($g:~S_3 \rightarrow S_3$) parametrized by
$g = \frac{1}{r} (x^0 {\bf 1} + i x^i \sigma^i)$ we get:
\be
\frac{1}{2\pi\aap} \int_{S_{3}} d\Sigma^{\mu\nu\lambda}
H_{\mu\nu\lambda} = \frac{Q}{2\pi\aap} \int_{S_3} tr \epsilon^{abc}
\left( (g^{-1} \pa_a g) (g^{-1} \pa_b g)  (g^{-1} \pa_c g) \right) =
2 \pi \frac{Q}{\aap}
\ee
and $\frac{Q}{\alpha'}$ has to be an integer \cite{witt1}, the topological
charge. The reason is the following. There are two possibilities to
continue the antisymmetric tensor contribution
to a torsion contribution, either to the outer side or to the inner
side of the world sheet. Because physically there is no difference we
have the upper consistency relation.
\section{Open string sector}
Now we want to look for a solution for the corresponding
\si model in open string sector. In this case there are two sets
of \bbb~functions which have to vanish in order to get a conformally
invariant
field theory \cite{dorn1,osbn}. On one hand these are again
the equations (6) corresponding to the background fields which couple at
inner points of the world sheet. These fields correspond to excitations of the
closed string. On the other hand we have in addition
\bbb~equations coming from fields which couple via the boundary
of the world sheet and  are the excitations of the open string theory:
the gauge field $A_{\mu}$ and a further tachyon field $T_2$. Crucial for the
further investigation is that the gauge field $A_{\mu}$ could only
influence the \bbb~functions (6) via higher genus
contributions \cite{call} (see also the discussion), i.e. at the lowest
genus the \bbb~functions (6)
are not changed by the boundary background fields. Therefore
the solution (2),(5) remains valid in our consideration. At this point
one has to note that the dilaton couples as well at inner points
of the world sheet as at the boundary. Consequently, the dilaton $\beta$
function {\em on} the boundary differs from the dilaton $\beta$ function
inside, e.g.~by additional gauge field contributions.
Later we will see that this does not matter for the solution (5).
But first of all we should discuss the gauge symmetry
of our theory (1). There are two gauge degrees of freedom: i) the usual
gauge of $A_{\mu}$ parameterized by a scalar field $\omega$ and
ii) the gauge of $B_{\mu\nu}$ parameterized by a vector filed
$\Lambda_{\mu}$
\be
\ba{l}
B_{\mu\nu} \rightarrow B_{\mu\nu} +  \, e \, \pa_{[\mu} \Lambda_{\nu]} \\
A_{\mu} \rightarrow A_{\mu} - \Lambda_{\mu} +
     \pa_{\mu} \omega ~.
\ea
\ee
The \bbb~functions depend only on the gauge invariant field strength
\be
F_{\mu\nu} = B_{\mu\nu} + e \, f_{\mu\nu} \qquad \qquad
( f_{\mu\nu}=\pa_{\mu}A_{\nu}-\pa_{\nu}A_{\mu})~.
\ee
Before we turn to the concrete form of the \bbb~functions let us do
some remarks concerning the perturbation theory with respect
to the gauge field coupling. Usually one considers the whole gauge field
coupling as an interaction part of the theory.
But it is possible to absorb the quadratic part of this interaction
by a modification of the boundary condition of the propagator.
This modified propagator is given by \cite{call}
\be
N_{\mu\nu} = -\frac{\aap}{2} \left(\eta_{\mu\nu} \log|z-z'|^2
            + \left(\frac{\eta-F}{\eta + F} \right)_{\mu\nu}
             \log(z-\bar{z}') + \left(\frac{\eta + F}{\eta-
             F} \right)_{\mu\nu} \log(\bar{z}-z') \right)~.
\ee
Using this propagator has the consequence that all gauge field
vertices contain at least
one derivative of the field strength $F$. If we neglect the higher
derivative terms of $F$ it is possible to obtain an exact
expression for the gauge field \bbb~function
\cite{call}. Let us now study the boundary \bbb~functions in detail.
Neglecting all higher derivatives in $F$ they are given by \cite{call,dorn1}
\be
\begin{array}{rcl}
\bbb^A _{\mu}&=&-\aap\left(\frac{1}{G-F^2}\right)^{\lambda\nu}
      D_{\lambda} F_{\nu\mu} + \frac{1}{2} M^{\nu} F_{\nu\mu} +  \frac{1}{2}
\left(\frac{F}{G-F^2}\right)^{\lambda\rho} H_{\lambda\rho}^{~~~\nu}F_{\nu\mu}
 -  \frac{1}{2 \, e} K_{\mu}~,\\
\bbb^{\phi}&=&\beta^{\phi} - \frac{1}{2} M^{\mu}\pa_{\mu}\phi~,\\
\bbb^{T_2}&=&\beta^{T_2} - T_2  + \frac{1}{2} M^{\mu}\pa_{\mu} T_2~.
\ea
\ee
We start with the discussion of the dilaton \bbb~function.
The Curci--Paffuti theorem \cite{curc,dorn1} which states that
the dilaton \bbb~function is constant if the \bbb~functions of the metric,
antisymmetric tensor and of the gauge field vanish is crucial
for this investigation.
Since the constant is the same for both dilaton \bbb~functions
\footnote{Because the dilaton
couples at inner points as well as at the boundary of the world sheet there
are two different dilaton \bbb~functions, e.g. only the boundary \bbb~function
depends on the gauge field (at the lowest genus).} \cite{osbn}
a vanishing inside causes a vanishing on the boundary. Therefore the
dilaton (5) is a zero of both \bbb~functions.
As in the closed string sector, $W_{\mu}$ and $K_{\mu}$ should
not contribute in the open string sector. Both quantities enter the
local Weyl anomaly as possible
total derivatives  and are absent in the global or integrated Weyl
anomaly \cite{tseyt2,dorn1}.  Because these terms are not relevant
for the inner
\bbb~functions (see the discussion before (9)) we assume that they also
not contribute to the boundary \bbb~functions. Hence we set again
\be
  W_{\mu} = K_{\mu} = 0 \qquad \mbox{or} \qquad M_{\mu} \equiv 2\aap
  \pa_{\mu}\phi ~.
\ee

Before we study the  gauge field \bbb~function one has to note that
in contrast to the closed string theory for open strings
the potential for the torsion  itself has physical
meaning similar to the Aharanov--Bohm effect.
In both theories only gauge invariant quantities enter the \bbb~functions.
While in the \si~model for closed strings only $H$ is a gauge invariant
quantity in open string theory there also appear the gauge invariant
quantity $F$. By the way for the gauge invariance of $F$ it is necessary
to couple a gauge field $A$. This antisymmetric tensor field, which is
also a possible potential for $H$, is entirely fixed by two equations
\be
\ba{c}
\pa_{[\mu} F_{\nu\lambda]}  \equiv \pa_{[\mu}B_{\nu\lambda]} =
   H_{\mu\nu\lambda} =
  \pm \epsilon_{\mu\nu\lambda\sigma}\pa^{\sigma} \log\frac{\sqrt{Q}}{r}~,\\
\bbb^A _{\mu} = 0 ~.
\ea
\ee
The first equation determines $F$ only up to a term $\sim
\pa_{[\mu} \Omega_{\nu]}$ and the second equation fixes $\Omega_{\nu}$.
The general solution of the first equation is given by
\be
\ba{c}
F_{\mu\nu} = -i \,\frac{3}{2}\, Q\,b_{[\mu}\,\pa_{\nu]}\chi +
\pa_{[\mu}\Omega_{\nu]} \\
\mbox{where:}\qquad b_{\mu} = J_{\mu}^{~\lambda}\pa_{\lambda}\log
   \frac{\sqrt{Q}}{r} \quad , \quad \chi = \log\frac{x\cdot p}{r} \quad ,
   \quad (J^2)_{\mu}^{~\nu} = -G_{\mu}^{~\nu}
\ea
\ee
and $p$ is an arbitrary eigenvector of the matrix $J$. There are six
possible matrices $J$
\be
\begin{array}{ll}
J_{1}^+ = \left(\begin{array}{cc} -i \sigma_2 & 0 \\
                                0 & i\sigma_2 \ea \right)~,&
  J_{1}^- = \left(\begin{array}{cc} -i\sigma_2 & 0 \\
                                0 & -i\sigma_2 \ea \right)~,\\
J_{2}^+ = \left(\begin{array}{cc} 0 & {\bf 1} \\
                                {\bf -1}& 0 \ea \right)~,&
  J_{2}^- = \left(\begin{array}{cc} 0 & \sigma_3  \\
                                -\sigma_3 & 0 \ea \right)~,\\
J_{3}^+ = \left(\begin{array}{cc} 0 & i\sigma_2  \\
                                i\sigma_2 & 0 \ea \right)~,&
  J_{3}^- = \left(\begin{array}{cc} 0 & -\sigma_1 \\
                                \sigma_1 & 0 \ea \right) ~.
\ea
\ee
These matrices are  discussed in \cite{strom} in the context
of (4,4) extended supersymmetric $\sigma$ model and they
are self dual ($J^+$) or anti-self-dual ($J^-$) respectively (see \cite{fre}).
All $J^+$ commute with all $J^-$ and both sets satisfy the algebra
\be
J^{\pm}_r J^{\pm}_{s} = - \delta_{r s} + \epsilon_{r s t} J^{\pm}_t ~.
\ee
In order to determine the vector $\Omega_{\mu}$ we insert $F$ in
$\bar{\beta}^A_{\mu}$  and find as solution
\be
\begin{array}{cc}
& -\aap \left(\frac{G}{G - F^2}\right)^{\lambda\nu}
 D_{\lambda} F_{\nu\mu} + \aap \pa^{\nu} \phi F_{\nu\mu} + \aap \frac{1}{2}
\left(\frac{F}{G-F^2}\right)^{\lambda\rho} H_{\lambda\rho}^{~~~\nu}F_{\nu\mu}
= 0\\
\mbox{if:} & \Omega_{\mu} = - i\, \frac{3}{4}\,Q\, b_{\mu} ~.
\end{array}
\ee
Let us note some general properties of $F$
\be
\ba{l}
(F^2)^2 = \frac{9}{4} F^2 \quad \rightarrow \quad
(G - F^2)^{-1} = G - \frac{4}{5} F^2 ~,\\
F \cdot x = F \cdot p = 0  ~,  \\
D_{(\nu} b_{\mu)} = 0 ~, \\
F \wedge F = 0 \qquad (\mbox{and also}~ R \wedge R = 0)
\ea
\ee
with $b_{\mu}$ is given in (19) and $p$ is an eigenvector of the matrix $J$.
The first projector property is useful in proving of (22) and
the second one will be used for the tachyon \bbb~function.
{}From the third equation follows that in addition to the covariantly
constant vector $\pa_{\mu}\log r$ a further Killing vector is given by
$b_{\mu}$ \footnote{ The three Killing vectors
$b_{\mu}^{(i)} = J_{\mu}^{(i) \nu}
\pa_{\nu} \log r$ (e.g. with the set of matrices $J^+$ in (20)) and the
covariantly constant vector $\pa_{\mu} \log r$  are an orthogonal
system of vectors.}.

Finally we have yet to fix the boundary tachyon $T_2$.
If we again assume that $T_2$ depends
on the radius $r$ only the corresponding \bbb~function is given by
\be
\bbb^{T_2} = - \aap D^2 T_2  - T_2 + \aap \pa_{\mu}\phi \pa^{\mu} T_2 = 0 ~.
\ee
This function looks very similar to the $\bar{\beta}^{T_1}$ function (11).
The only differences are the factors in front of the first and second
term. The first term has to be twice the term in (11) because the
divergence {\em on} the boundary is twice than
inside \cite{dorn}.
In the discussion of the $\bar{\beta}^{T_1}$ function we pointed
out that curvature and torsion terms have to drop out because the
covariantly constant curvature
and torsion are ``transverse'' to any covariant derivative of the tachyon.
The same arguments are true for $T_2$. In addition to the curvature
and torsion terms all $F_{\mu\nu}$ dependence drops out because
\footnote{We neglect all higher derivatives of $F_{\mu\nu}$.}:
$F_{\mu\nu} x^{\nu} = 0$ (see (23)) and $x^{\nu} D_{\lambda} F_{\nu\mu}=0$.
In terms of the dilaton (5)
$ \phi = q \log\frac{r}{r_0}$  we find as solution
\be
T_2 \sim r^{\alpha_{\pm}} \qquad , \qquad \alpha_{\pm} = \frac{1}{2}
   \sqrt{\frac{k}{6}} \left(\sqrt{25 - \frac{3k}{k+2}} \pm \sqrt{1 -
   \frac{3k}{k+2}}\right) ~.
\ee
Thus both tachyons have the same functional structure
but: $\alpha_{\pm}^{closed} = 2 \alpha_{\pm}^{open}$. Of course, in terms of
$\sin$ or $\cos$ function it is possible to get two real solutions for
all $k$ like (5).
\section{Discussion}
Summarizing all results the background fields
for which the Weyl anomaly vanish are given by
\be
\ba{l}
G_{\mu\nu}=\frac{Q}{r^2}\delta_{\mu\nu} \quad , \quad
H_{\mu\nu\lambda}=\pm\epsilon_{\mu\nu\lambda}^{~~~~\sigma}
  \partial_{\sigma}\log\frac{\sqrt{Q}}{r} \qquad  \mu,\nu = 0,1,2,3\\
F_{\mu\nu} \equiv B_{\mu\nu} + e\,f_{\mu\nu} = - i\,\frac{3}{2}\,Q
  \left( b_{[\mu} \pa_{\nu]} \chi + \frac{1}{2} \partial_{[\mu} b_{\nu]}
  \right)\\
\phi = \phi_0 + q \log r \\
T_1 \sim r^{\alpha_{\pm}} \quad , \quad
T_2 \sim r^{\frac{1}{2}\alpha_{\pm}} \quad , \quad \alpha_{\pm} =
   \sqrt{\frac{k}{6}} \left(\sqrt{25 - \frac{3k}{k+2}} \pm \sqrt{1 -
   \frac{3k}{k+2}}\right)
\ea
\ee
where $b_{\mu}$ and $\chi$ are defined in (19).
The metric could be understood as an approximation of a semi-wormhole
near the singularity \cite{strom}. This semi-wormhole is given by a
conformally flat metric with the conformal factor: $\sim C + \frac{Q}{r^2}$.
Because the geometry of the space time in this limit is a
cylinder with a three-sphere as cross--section
\be
ds^2 = \frac{Q}{r^2}(dr^2 + r^2 d\Omega_3^2) = dt^2 + Q d\Omega_3^2
\quad , \quad t = \sqrt{Q}\,\log r
\ee
it is also possible to regard the space time
as a Robertson--Walker universe with the world radius $Q$
\cite{elis,behr}. After this transformation we get a dilaton which
depends linearly on the time with $q$ as a background charge.
The tachyon fields differ only by a factor 2 in front of $\alpha_\pm$
and are in general oscillating fields of the time $t$.
Only for $k=1$ we get after the
transformation (27) an exponential time-like tachyon. An additional
spatial dependence of the tachyon for closed strings is discussed in
\cite{behr}.

Now we determine the gauge field $A_{\mu}$ explicitly.
The \bbb~functions define only $F_{\mu\nu} \equiv B_{\mu\nu} +
e \, f_{\mu\nu} = $$ - i\,\frac{3}{2}\, Q (b_{[\mu}
\pa_{\nu]} \chi + \frac{1}{2}\,\pa_{[\mu} b_{\nu]} )$ but not the
field strength $f_{\mu\nu} = \pa_{[\mu} A_{\nu]}$. An obvious
choice would be: $A_{\mu} \sim b_{\mu}$. But $F$ can also be written
as $F_{\mu\nu} \sim \chi \pa_{[\mu} b_{\nu]} +
\pa_{[\mu}(1 - \chi)b_{\nu]}$ which would yield: $A_{\mu} \sim
(1 - \chi) b_{\mu}$. Both  gauge fields are equivalent from the
\si model point of view. They differ only by a special gauge transformation.
In order to decide between both fields we have to impose a gauge fixing
condition. If we take, e.g., the Lorentz gauge condition for the more general
gauge field $A_{\mu} = (c \chi + d)b_{\mu}$ ($c, d$ are some constants)
we find
\be
D_{\mu} A^{\mu} = D_{\mu} ( c \chi + d ) b^{\mu} = c \, b^{\mu}
\pa_{\mu}\chi = 0 \quad \leftrightarrow \quad c = 0 ~.
\ee
Hence a possible gauge field and antisymmetric tensor are given by
\be
A_{\mu} = -i \,\frac{3}{4}\,\frac{Q}{e}\, b_{\mu} \qquad
\mbox{and} \qquad B_{\mu\nu} = - e \, A_{[\mu} \pa_{\nu]} \chi ~.
\ee
{}From (23) it follows that for this gauge fixing $A_{\mu}$ is just
a Killing vector field.

We should remark that a priori $F_{\mu\nu}$ is not well defined
everywhere. Because $H = dF$ and $\int H \neq 0$ we have to expect a
Dirac singularity
in $F_{\mu\nu}$. Indeed, performing the partial derivatives in $F$
yields
\be
F_{\mu\nu} = -i\,\frac{3}{2}\, Q  \left( \frac{1}{r^2} J_{\mu\nu} +
\frac{1}{(xp)}(b_{\mu}    p_{\nu} -  b_{\nu} p_{\mu})\right)
\ee
and we find a Dirac singularity for $|xp| = 0$ (note: the vectors
$p_{\mu}$ are complex and we have to take the real part only).
Therefore, we have
to be careful in inserting the antisymmetric tensor in the \si model (1).
This can be done by seperating the manifold $\Sigma$ in two
parts ($\Sigma^+$ ,
$\Sigma^-$) and choosing for each part an eigenvector which does not cause
a singularity in this area\footnote{For $J_1^+$ the eigenvectors are
given by: $(1,i,0,0) , (0,0,1,i)$ and the complex conjugate vectors.}.
Since $F_{\mu\nu}$ satisfies
\be
\ba{l}
^{\ast} d\, ^{\ast} F = 0 \\
^{\ast} d\, F = j^{M} = Q\, d \log r ~.
\ea
\ee
we see that $F_{\mu\nu}$ fulfills the equation of motion for a monopole field.
To get the magnetic charge we have to perform the integration
\be
\int_{S^2} F = \int_{M_3} H = Q
\ee
where $M_3$ is a spatial 3-manifold with the boundary $S_2$.
The magnetic current is conserved ($D^{\mu}
j^{M}_{\mu} = 0$) and the quantization of the magnetic charge $Q$ follows
from (12). If we again take $t=\sqrt{Q} \log r$ as time,
the spatial part of the torsion $H$ is given by: $H_{ijk} = Q \epsilon_{ijk}$.
With $H_{ijk} = \pa_{[i}F_{jk]}$ one finds \cite{brat}:
$F_{ij} \sim \epsilon_{ijk} x^k$ and therefore we
get the expected radial magnetic field: $B^i = \epsilon^{ijk} F_{jk}
\sim x^{i}$.

\vsf

Before we discuss the effective action corresponding to these
background fields let us consider the \bbb~function once more.
It is possible to express these functions in terms of
Lie derivatives \cite{tseyt2,curc}
\be
\begin{array}{ll}
\bbb^i = \beta^i + \sigma^i \qquad , & \sigma^{G}_{\mu\nu} = {\cal L}_{M}
                                   G_{\mu\nu} ~,\\
      & \sigma^{B}_{\mu\nu} = {\cal L}_{M} B_{\mu\nu} - \pa_{[\mu}
                       (M \cdot  B)_{\mu]} ~, \\
      & \sigma^{A}_{\mu} = {\cal L}_{M} A_{\mu} +
                      \frac{1}{2\pi\aap}(M \cdot B)_{\mu}
                     -       \pa_{\mu} (M \cdot  A) ~,   \\
      & \sigma^{T_1} = {\cal L}_{M} T_1 - 2 T_1 ~,\\
      & \sigma^{T_2} = {\cal L}_{M} T_2 -  T_2 ~,\\
      & \sigma^{\phi} = {\cal L}_{M} \phi
\end{array}
\ee
where the vector $M_{\mu} = 2 \aap \pa_{\mu}\phi$ yields the
direction of the Lie derivative. If we insert the fields
(26), (29) we find that the Lie derivative vanish for the metric,
antisymmetric tensor and for the gauge field. Furthermore, one finds:
$(M \cdot B)_{\mu} = (M \cdot A) = 0$ and hence $\sigma^{G,B,A}=0$.
This has the consequence that for these background fields the
renormalization group $\beta$ function itself vanish
and thus all counterterms. On the other side $\sigma^{T_{1,2}}$
and $\sigma^{\phi}$ do not vanish for our solution, i.e.~the
tachyons as well as the dilaton have to be renormalized.
Although the dilaton is a free field (after the transformation (27) it is
linear in time) we have an additive
renormalization of $\phi$. The vanishing of the dilaton \bbb~function
is equivalent to the vanishing of the total central charge where
the $\sigma$ coefficient contains just the contribution of the background
charge $q$.

Let us shortly discuss the mass of the tachyons.
We define the mass of the tachyon via the equation of motion, i.e.~the
\bbb~equations (11) and (24). In order to read off the mass we have to
decouple the tachyon field from the dilaton. This can be done in terms of
the transformation $T_1 = e^{\phi} \tilde{T_1}$ and we get after using
the dilaton \bbb~equation for $\tilde{T_1}$
\begin{equation}
-\frac{1}{2} \alpha' D^2 \tilde{T} - \frac{1}{12}
\left(\frac{3k}{k+2}-1\right) \tilde{T} = 0 ~.
\end{equation}
For $T_2$ one obtains a similar expression with one half of the mass.
Thus the mass of both tachyons
are proportional to: $(1 - \frac{3k}{k+2})$ and we get
massless tachyons only if $k = 1$ or $c_{wzw}=1$ \cite{behr}.
In this case: $\alpha_+ = \alpha_- = q = 2$ and after the transformation (27)
the massless solutions for $T_{1,2}$ are given by
\be
\ba{l}
T_1^{(1)} \sim e^{2 t} \quad , \quad T_1^{(2)} \sim \, t \,e^{2 t}\\
T_2^{(1)} \sim e^{t} \quad , \quad T_2^{(2)} \sim \, t \,e^{t} ~.
\ea
\ee
Here we should remark that all results we were computed  in
an $\aap$ expansion. On the other hand in a weak field expansion for
the tachyon one has to take into account additional divergencies,
e.g.~$T^2$ terms. If one has an exponential
tachyon as in the Liouville theory one can avoid divergencies like these
by an appropriate continuation in the momentum like in the computation
of Shapiro--Virasoro amplitudes. But this procedure does not work for
oscillating tachyons. Applying this statement to our solution
means that only massless tachyons (34) would yield an exact conformal
field theory namely the Liouville theory. Therefore the incorporation
of non-perturbative divergencies would restrict us to $k=1$.

\vsf

Finally, we discuss the effective action yielding the fields (26).
In the open string sector the gauge field \bbb~function follows from
the Born-Infeld action \cite{born}
\be
\begin{array}{rcl}
S_{open}&=&- \kappa \int d^4 x\sqrt{det(G + F)}e^{- \phi} = -\kappa
  \int d^4 x \, \sqrt{ det G }\, e^{-\phi}\, \sqrt{ det(1 + G^{-1} F) } = \\
    &=&- \kappa \int d^4 x \,
  \sqrt{ det G }\, e^{-\phi}\, \sqrt{ 1 + \frac{1}{2} F^2 +
  \frac{1}{64} (F \wedge F)^2 }
\end{array}
\ee
where $\kappa$ has to be positive in order to couple the gauge field to
gravity with the right sign.
In the closed string sector we have the known effective action \cite{call1}
\be
S_{closed}=\int \sqrt{ G } \, e^{-2 \phi}\left[-\frac{2}{3}(26-c)
  + \aap ( R + 4(\pa\phi)^2 - \frac{1}{12} H^2 ) +
  {\cal O}(\alpha'^{2}) \right]
\ee
where $c$ in our model is given by: $c = 1 + \frac{3k}{k+2}$.
In the flat limit ($k\rightarrow \infty$) we get $c=4$, the number
of space time dimensions.
One should note here that the incorporation of the tachyons is not
straightforward. The reason is the following. In our consideration
we consider the \si model strictly in an $\aap$ expansion. In this
expansion the tachyons do not influence, e.g.~the metric, because
there are no counterterms to the metric. Only in a weak field
expansion for the tachyon the metric \bbb~function contains tachyon
contributions \cite{tseyt1}.
This has the consequence that we can not simply incorporate the tachyon
in the effective action in a covariant manner as a matter field.
Therefore we want to neglect the tachyon fields in this consideration.
The total effective action is then given by the sum \cite{call}
\be
S_{eff}=S_{closed} + S_{open}
\ee
In the equations of motion following from this effective action are
additional terms, e.g. to the metric \bbb~function (6) we have to add
the contribution
\be
\sim \kappa e^{\phi} \sqrt{det(1+G^{-1} F)}\left(\frac{G+F^2}
{G-F^2}\right)_{\mu\nu} ~.
\ee
Loop calculations show that terms like this can be interpreted as
loop-correction to the metric \bbb~ function \cite{call}. These
contributions correspond to a gauge field couplings via small holes
in the world sheet, i.e.~higher genus contributions.
Performing the integration of the radius from the
hole causes this divergent term.

In order to decouple the dilaton from the graviton it is common
to perform the following Weyl rescalling \cite{tseyt2}
\be
G_{\mu\nu} \rightarrow \tilde{G}_{\mu\nu} = e^{-2\phi} G_{\mu\nu}
\ee
and one obtains the action \cite{call1}
\be
\begin{array}{c}
S=\int \sqrt{ \tilde{G} } \left[\aap \tilde{R} -\aap 2 (\pa\phi)^2
   - \aap \frac{1}{12} H^2 e^{-4 \phi}  + \frac{2}{3}(26-c) e^{2 \phi}
    -\right. \\
 \left. -  \kappa e^{3 \phi}\, \sqrt{ 1 + \frac{1}{2} \tilde{F}^2 +
  \frac{1}{64} (\tilde{F} \wedge \tilde{F})^2 } + ... \right]
\end{array}
\ee
with $\tilde{F}_{\mu\nu} = e^{- 2 \phi} F_{\mu\nu}$. The equation
of motion e.g.~for the metric is now given by
\be
\tilde{R}_{\mu\nu} -\frac{1}{2} \tilde{G}_{\mu\nu} \tilde{R} =
  T_{\mu\nu}^{matter}
\ee
with the energy momentum tensor
\begin{equation}
\begin{array}{rcl}
T_{\mu\nu}^{matter}&=& \frac{1}{3 \aap}(26-c)\, e^{2 \phi}\,
 \tilde{G}_{\mu\nu}
 + \frac{1}{4} \left( H^2_{\mu\nu} - \frac{1}{6} \tilde{G}_{\mu\nu}
 H^2 \right)  e^{-2 \phi}+\\
& & +  \left( \partial_{\mu} \phi \partial_{\nu} \phi
 - \frac{1}{2} (\partial \phi )^2 \tilde{G}_{\mu\nu} \right) - \kappa
  e^{3 \phi}\, \sqrt{ 1 + \frac{1}{2} \tilde{F}^2 +
  \frac{1}{64} (\tilde{F} \wedge \tilde{F})^2 } \left(\frac{\tilde{G}}
  {\tilde{G} - \tilde{F}^2} \right)_{\mu\nu} ~.
\end{array}
\end{equation}
Because the field strength
contribution correspond to a loop-correction to $\bar{\beta}^{G}$
which we have neglected, our solution fulfills this metric equation
of motion only for $\kappa=0$. If we take our solution (26)
and perform the Weyl transformation (40),
we find for the metric
\be
\begin{array}{rcl}
\tilde{d s}^2&=&e^{-2 \phi} \frac{Q}{r^2}(d r ^2 + d\Omega_3^2)\\
      &=&d\tau^2 + \frac{q^2 \tau^2}{[1+\frac{1}{4}\bar{r}^2]^2}
   \left( (dx^1)^2 + (dx^2)^2 + (dx^3)^2 \right)
\end{array}
\ee
where: $q^2 = \frac{k}{6}(25-\frac{3k}{k+2})$ and
$\bar{r}=(x^1)^2+(x^2)^2+(x^3)^2$. This is just the Robertson--Walker
metric for a linear expanding universe with the world radius:
$K^2=q^2 \tau^2$. The possible values of $q$
are restricted if we include non-perturbative contributions for
the tachyons (see the discussion after (35)). In this case the
tachyons have to be massless ($k=1$) and therewith $q=2$.
Furthermore, if we insert our solution in (43) and after
a time reparametrization we get for the
energy momentum tensor (for $\kappa=0$)
\be
T_{\mu\nu}^{matter} = (\frac{1}{2 q^2} +2 ) \, \pa_{\mu}\log\tau \,\pa_{\nu}
\log\tau \, - \, \frac{1}{2}(\frac{1}{2 q^2} - 2) \frac{1}{\tau^2}\,
 \tilde{G}_{\mu\nu}
\ee
which correspond to an ideal liquid with the energy density:
$\mu= (\frac{1}{4}+2q^2) \frac{1}{q^2 \tau^2}$ and the isotropic
pressure: $p=(q^2-\frac{1}{4}) \frac{1}{q^2 \tau^2}$.
\section{Conclusion}
In this paper we have firstly summarized the known results for
a conformally exact \si model in closed string theory
\cite{khur,elis,behr}: a combination of the SU(2) WZW model with the
Liouville theory. In addition, we have discussed this model from the
\si model point
of view as solution of the \bbb~equations to all orders in $\aap$.
In the second part we generalized these results to an
open string theory where we coupled an abelian gauge field and a further
tachyon via the boundary of the world sheet. Since we have neglected
loop corrections (higher genus), the solution of the closed string model
remained unchanged  and we had yet to look for a solution for the gauge field
and boundary tachyon  $\bbb$ equation. Because we considered in this \si
model an antisymmetric tensor as well as a gauge field coupling, we had
to take into account two different gauge transformations.
A gauge invariant field strength was given by the sum of the
antisymmetric tensor and the usual field strength. Assuming that the tachyon
depends only on the radius and neglecting all contributions
which are quadratic or higher in the derivatives of the field strength
we got for the open string sector an exact
solution too (see (26)). This solution fulfills a monopole equation
with a quantized
magnetic charge and after a gauge fixing the corresponding gauge field
was just given by a Killing vector field of this theory.
The investigation of the tachyons showed that both tachyons have
quantized masses. This quantization is given by the central charge
of the SU(2) WZW model and both tachyons are massless if $c_{wzw}=1$.
In addition, only the massless tachyons are not
oscillating in the time and yield a conformal field theory
also after incorporation of non-perturbative contributions.
Finally, we have discussed the effective action where the
corresponding gauge field equation of motion follows from a
Born-Infeld action. The space time can be interpreted as a
linear expanding closed universe.

\vspace{8mm} \noindent
{\large\bf Acknowledgments} \vspace{3mm} \newline \noindent
We would like to thank H.~Dorn, S.~F\"orste, J.~Schnittger and G.~Weigt
for helpful discussions.

\renewcommand{\arraystretch}{1.0}

\end{document}